\documentclass[runningheads]{llncs}
\usepackage{graphicx}
\usepackage{amsfonts}
\usepackage{mathtools}
\usepackage{amssymb}
\usepackage{diagbox}
\newcommand\Tstrut{\rule{0pt}{2.4ex}}         
\newcommand\Bstrut{\rule[-0.9ex]{0pt}{0pt}}   

\begin{document}
\title{Transformer-based Dual-domain Network for Few-view Dedicated Cardiac SPECT Image Reconstructions}
%
%
\author{Huidong Xie$^1$, Bo Zhou$^1$, Xiongchao Chen$^1$, Xueqi Guo$^1$, Stephanie Thorn$^1$, Yi-Hwa Liu$^1$, Ge Wang$^2$, Albert Sinusas$^1$, Chi Liu$^1$}

\institute{$^1$Yale University, New Haven, CT 06511, USA \\
$^2$Rensselaer Polytechnic Institute, Troy, NY 12180, USA\\
\email{\{huidong.xie, chi.liu\}@yale.edu}}
%

\authorrunning{Xie et al.}

\maketitle              
\begin{abstract}
Cardiovascular disease (CVD) is the leading cause of death worldwide, and myocardial perfusion imaging using SPECT has been widely used in the diagnosis of CVDs. The GE 530/570c dedicated cardiac SPECT scanners adopt a stationary geometry to simultaneously acquire 19 projections to increase sensitivity and achieve dynamic imaging. However, the limited amount of angular sampling negatively affects image quality. Deep learning methods can be implemented to produce higher-quality images from stationary data. This is essentially a few-view imaging problem. In this work, we propose a novel 3D transformer-based dual-domain network, called TIP-Net, for high-quality 3D cardiac SPECT image reconstructions. Our method aims to first reconstruct 3D cardiac SPECT images directly from projection data without the iterative reconstruction process by proposing a customized projection-to-image domain transformer. Then, given its reconstruction output and the original few-view reconstruction, we further refine the reconstruction using an image-domain reconstruction network. Validated by cardiac catheterization images, diagnostic interpretations from nuclear cardiologists, and defect size quantified by an FDA 510(k)-cleared clinical software, our method produced images with higher cardiac defect contrast on human studies compared with previous baseline methods, potentially enabling high-quality defect visualization using stationary few-view dedicated cardiac SPECT scanners.

\end{abstract}
\section{Introduction}
The GE Discovery NM Alcyone 530c/570c \cite{bocher_fast_2010} are dedicated cardiac SPECT systems with 19 cadmium zinc telluride (CZT) detector modules designed for stationary imaging. Limited amount of angular sampling on scanners of this type could affect image quality. Due to the unique geometry of Alcyone scanners, the centers of FOV vary at different angular positions. Hence, unlike CT scanners, there is no straightforward method to combine projections at different positions on Alcyone scanners. Xie \textit{et al.} \cite{xie_increasing_2022} proposed to incorporate the displacements between centers of FOV and the rotation angles into the iterative method for multi-angle reconstructions with registration steps within each iteration. Despite its promising results, acquiring multi-angle projections on this scanner is time-consuming and inconvenient in reality. Rotating the detectors also limits the capability of dynamic imaging. Thus, it is desirable to obtain high-quality rotation-based reconstruction directly from the stationary SPECT projection data. This is essentially a few-view reconstruction problem.

Previous works have attempted to address this problem by using deep-learning-based image-to-image networks. Xie \textit{et al.} proposed a 3D U-net-like network to directly synthesize dense-view images from few-view counterparts \cite{xie_increasing_2022}. Since convolutional networks have limited receptive fields due to small kernel size, Xie \textit{et al.} further improved their method with a transformer-based image-to-image network for SPECT reconstruction \cite{xie_deep_2022}. Despite their promising reconstruction results, these methods use MLEM (maximum likelihood expectation maximization) reconstruction from one-angle few-view acquisition data as network input. The initial MLEM reconstruction may contain severe image artifacts with important image features lost during the iterative reconstruction process, thus would be challenging to be recovered with image-based methods. Learning to reconstruct images directly from the projection data could lead to improved quality.

There are a few previous studies proposed to learn the mapping between raw data and images. AUTOMAP \cite{zhu_image_2018} utilized fully-connected layers to learn the inverse Fourier transform between k-space data and the corresponding MRI images. While such a technique could be theoretically adapted to other imaging modalities, using a similar approach would require a significant amount of trainable parameters, and thus is infeasible for 3D data. W\"{u}rfl \textit{et al.} \cite{wurfl_deep_2018} proposed a back-projection operator to link projections and images to reduce memory burden for CT. However, their back-projection process is not learnable. There are also recent works \cite{he_radon_2020,li_learning_2019,xie_deep_2020} that tried to incorporate the embedded imaging physics and geometry of CT scanners into the neural networks to reduce redundant trainable parameters for domain mapping. However, these networks are unable to be generalized to other imaging modalities due to different physical properties. Moreover, these methods are hard to be extended to 3D cases because of the geometric complexity and additional memory/computational burden.

Here, we propose a novel 3D Transformer-based Dual-domain (projection \& image) Network (TIP-Net) to address these challenges. The proposed method reconstructs high-quality few-view cardiac SPECT using a two-stage process. First, we develop a 3D projection-to-image transformer reconstruction network that directly reconstructs 3D images from the projection data. In the second stage, this intermediate reconstruction is combined with the original few-view reconstruction for further refinement, using an image-domain reconstruction network. Validated on physical phantoms, porcine, and human studies acquired on GE Alcyone 570c SPECT/CT scanners, TIP-Net demonstrated superior performance than previous baseline methods. Validated by cardiac catheterization (Cath) images, diagnostic results from nuclear cardiologists, and cardiac defect quantified by an FDA-510(k)-cleared software, we also show that TIP-Net produced images with higher resolution and cardiac defect contrast on human studies, as compared to previous baselines. Our method could be a clinically useful tool to improve cardiac SPECT imaging.


\section{Methodology}
Following the acquisition protocol described by Xie \textit{et al.} \cite{xie_increasing_2022}, we acquired a total of eight porcine and two physical phantom studies prospectively with four angles of projections. The training target in this work is four-angle data with 76 ($19\times4$) projections and the corresponding input is one-angle data with 19 projections. Twenty clinical anonymized $^{99m}$Tc-tetrofosmin SPECT human studies were retrospectively included for evaluation. Since only one-angle data was available for the human studies, Cath images and clinical interpretations from nuclear cardiologists were used to determine the presence/absence of true cardiac defects and to assess the images reconstructed by different methods. The use of animal and human studies was approved by the Institutional Animal Care \& Use Committee (IACUC) of Yale University.

\subsection{Network Structure}
\begin{figure*}[!ht]
\centering
\makebox[\textwidth][c]{\includegraphics[width=0.8\textwidth]{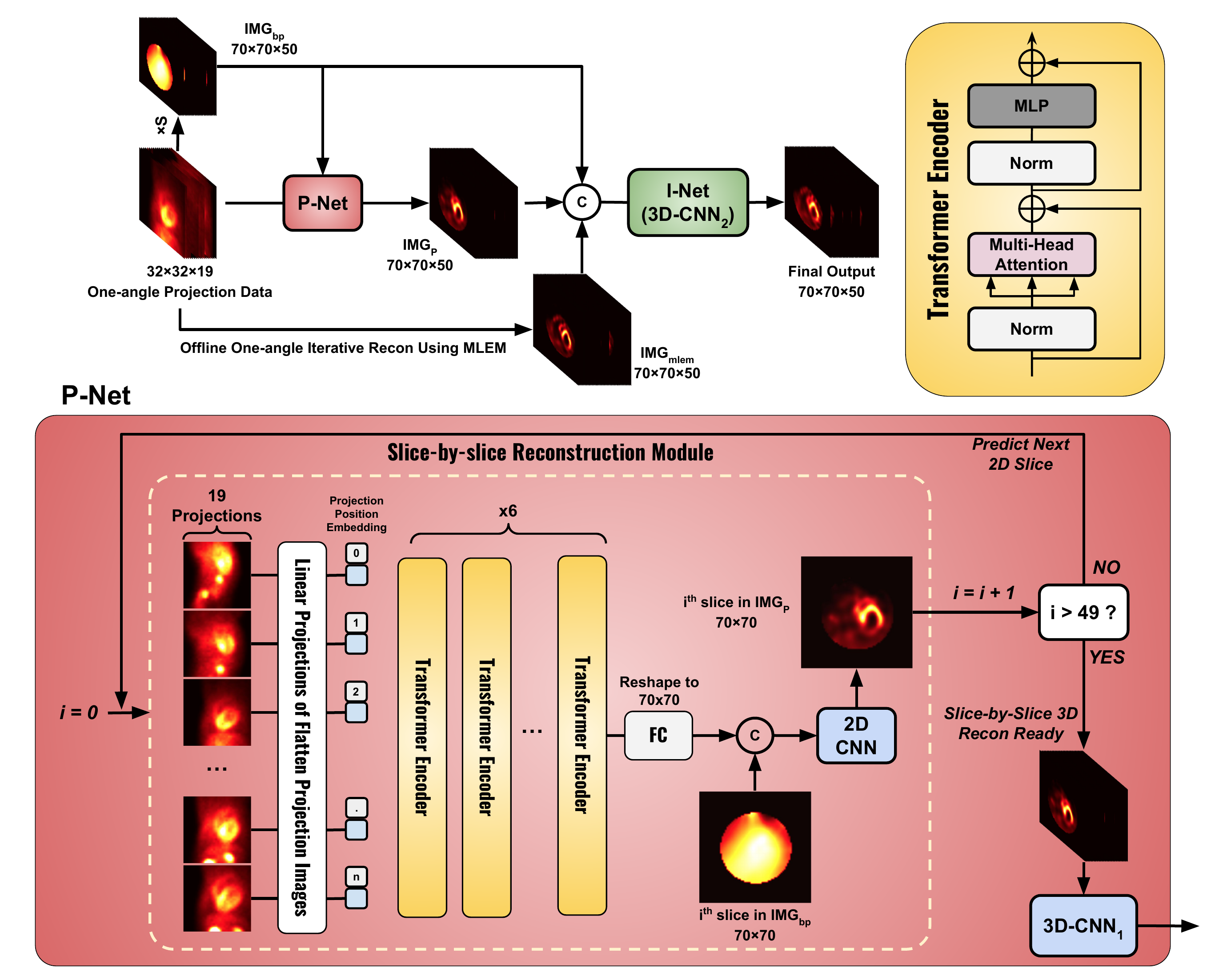}}
\centering
\caption{Overall TIP-Net structure. The TIP-Net is divided into P-net and I-net. The back-projected image volume was generated by multiplying the 3D projections with the system matrix. FC layer: fully-connected layer. Both 3D-$\mathrm{CNN_1}$ and 3D-$\mathrm{CNN_2}$ share the same structure but with different trainable parameters.}
\label{fig:1}
\end{figure*}

The overall network structure is presented in Fig. \ref{fig:1}. TIP-Net is divided into two parts. The transformer-based \cite{dosovitskiy_image_2021} projection-net (P-net) aims at reconstructing 3D SPECT volumes directly from 3D projections obtained from the scanner. Information from the system matrix ($\mathrm{IMG_{bp}}$) is also incorporated in P-net as prior knowledge for image reconstruction. $\mathrm{IMG_{bp}}$ is obtained by multiplying the system matrix $\mathrm{S}\in \mathbb{R}^{19,456\times245,000}$ with the projection data. The outputs from P-net ($\mathrm{IMG_{p}}$) serve as an input for image-net (I-net). 

To reconstruct $\mathrm{IMG_{p}}$ in P-net, we may simply project the output from the Transformer block to the size of 3D reconstructed volume (i.e., $70\times70\times50$). However, such implementation requires a significant amount of GPU memory. To alleviate the memory burden, we proposed to learn the 3D reconstruction in a slice-by-slice manner. The output from the Transformer block is projected to a single slice ($70\times70$), and the whole Transformer network (red rectangular in Fig.\ref{fig:1}) is looped 50 times to produce a whole 3D volume ($70\times70\times50$). Since different slice has different detector sensitivity, all the 50 loops use different trainable parameters and the $i^{th}$ loop aims to reconstruct the $i^{th}$ slice in the 3D volume. Within each loop, the Transformer block takes the entire 3D projections as input to reconstruct the $i^{th}$ slice in the SPECT volume. With the self-attention mechanism, all 50 loops can observe the entire 3D projections for reconstruction. The $70\times70$ slice is then combined with $\mathrm{IMG_{bp}}$ (only at $i^{th}$ slice), and the resized 3D projection data. The resulting feature maps ($70\times70\times21$) are fed into a shallow 2D CNN to produce a reconstructed slice at the $i^{th}$ loop ($70\times70\times1$), which is expected to be the $i^{th}$ slice in the SPECT volume.

I-net takes images reconstructed by P-net ($\mathrm{IMG_{p}}$), prior knowledge from the system matrix ($\mathrm{IMG_{bp}}$), and images reconstructed by MLEM ($\mathrm{IMG_{mlem}}$) using one-angle data for final reconstructions. With such a design, the network can combine information from both domains for potentially better reconstructions compared with methods that only consider single-domain data.

Both $\mathrm{3D\text{-}CNN_1}$ and $\mathrm{3D\text{-}CNN_2}$ use the same network structure as the network proposed in \cite{xie_increasing_2022}, but with different trainable parameters.

\subsection{Optimization, Training, and Testing}

In this work, the TIP-Net was trained using a Wasserstein Generative Adversarial Network (WGAN) with gradient penalty \cite{NIPS2017_892c3b1c}. A typical WGAN contains 2 separate networks, one generator ($G$) and a discriminator ($D$). In this work, the network $G$ is the TIP-Net depicted in Fig.\ref{fig:1}. Formulated below, the overall objective function for network $G$ includes SSIM, MAE, and Adversarial loss.

\begin{equation}
\begin{split}
\underset{{\theta}_G}{\min}\,L = \bigg\{\ell(G(I_{\mathrm{one}}), I_{\mathrm{four}})+\lambda_a\ell(\mathrm{P_{net}}(I_{\mathrm{one}}), I_{\mathrm{four}})
- \underbrace{\lambda_b\,\mathbb{E}_{I_{\mathrm{one}}}\left[D(G({I}_{\mathrm{one}}))\right]}_{\mathrm{Adversarial\ Loss}}\bigg\},
 \label{eqn:1}
 \end{split}
\end{equation}
where $I_{\mathrm{one}}$ and $I_{\mathrm{four}}$ denote images reconstructed by one-angle data and four-angle data respectively using MLEM. $\theta_G$ represents trainable parameters of network $G$. $\mathrm{P_{net}}(I_{\mathrm{one}})$ represents the output of the P-net ($\mathrm{IMG_p}$). The function $\ell$ is formulated as below:

\begin{equation}
\begin{split}
\ell(X,Y) =
 \ell_{\mathrm{MAE}}(X, Y) + \lambda_c\,  \ell_{\mathrm{SSIM}}(X, Y) +  \lambda_d\, \ell_{\mathrm{MAE}}(\mathrm{SO}(X), \mathrm{SO}(Y)) ,
 \label{eqn:2}
 \end{split}
\end{equation}
where $X$ and $Y$ represent two image volumes used for calculations. $\ell_{\mathrm{MAE}}$ and $\ell_{\mathrm{SSIM}}$ represent MAE and SSIM loss functions, respectively. The Sobel operator ($\mathrm{SO}$) was used to obtain edge images, and the MAE between them was included as the loss function. $\lambda_a=0.1$, $\lambda_b=0.005$, $\lambda_c=0.8$, and $\lambda_d=0.1$ were fine-tuned experimentally. Network $D$ shares the same structure as that proposed in \cite{xie_increasing_2022}.

The Adam method \cite{DBLP:journals/corr/KingmaB14} was used for optimizations. The parameters were initialized using the Xavier method \cite{pmlr-v9-glorot10a}. 250 volumes of simulated 4D extended cardiac-torso (XCAT) phantoms \cite{segars_4d_2010} were used for network pre-training. Leave-one-out testing process was used to obtain testing results for all the studies.

\subsection{Evaluations}
Reconstructed images were quantitatively evaluated using SSIM, RMSE, and PSNR. Myocardium-to-blood pool (MBP) ratios were also included for evaluation. For myocardial perfusion images, higher ratios are favorable and typically represent higher image resolution. Two ablated networks were trained and used for comparison. One network, denoted as $\mathrm{3D\text{-}CNN}$, shared the same structure as either $\mathrm{3D\text{-}CNN_1}$ or $\mathrm{3D\text{-}CNN_2}$ but only with $\mathrm{IMG_{mlem}}$ as input. The other network, denoted as $\mathrm{Dual\text{-}3D\text{-}CNN}$, used the same structure as the TIP-Net outlined in Fig.\ref{fig:1} but without any projection-related inputs. Compared with these two ablated networks, we could demonstrate the effectiveness of the proposed projection-to-image module in the TIP-Net. We also compared the TIP-Net with another transformer-based network (SSTrans-3D) \cite{xie_deep_2022}. Since SSTrans-3D only considers image-domain data, comparing TIP-Net with SSTrans-3D could demonstrate the effectiveness of the transformer-based network for processing projection-domain information.

To compare the performance of cardiac defect quantifications, we used the FDA 510(k)-cleared Wackers-Liu Circumferential Quantification (WLCQ) software \cite{liu_quantification_1999} to calculate the myocardial perfusion defect size (DS). For studies without cardiac defect, we should expect lower measured defect size as the uniformity of myocardium improves. For studies with cardiac defects, we should expect higher measured defect size as defect contrast improves.

Cath images and cardiac polar maps are also presented. Cath is an invasive imaging technique used to determine the presence/absence of obstructive lesions that results in cardiac defects. We consider Cath as the gold standard for the defect information in human studies. The polar map is a 2D representation of the 3D volume of the left ventricle. All the metrics were calculated based on the entire 3D volumes. All the clinical descriptions of cardiac defects in this paper were confirmed by nuclear cardiologists based on SPECT images, polar maps, WLCQ quantification, and Cath images (if available).

\section{Results}
\subsection{Porcine results}
Results from one sample porcine study are presented in Fig. S1. This pig had a large post-occlusion defect created by inflating an angioplasty balloon in the left anterior descending artery. As pointed out by the green arrows in Fig. S1, all neural networks improved the defect contrast with higher MBP ratios compared with the one-angle images. The TIP-Net results were better than other network results in terms of defect contrast based on defect size measured by WLCQ. TIP-Net also produced images with clearer right ventricles, as demonstrated by the Full-width at Half Maximum (FHMW) values presented in Fig. S1.

Quantitative results for 8 porcine and 2 physical phantom studies are included in Table \ref{tab1}. Based on paired t-tests, all the network results are statistically better than one-angle results ($p<0.001$), and the TIP-Net results are statistically better than all the other three network results ($p<0.05$). The average MBP ratios shown in Table \ref{tab1} indicate that the proposed TIP-Net produced images with higher resolution compared with image-based networks.

Average defect size measurements for porcine and physical phantom studies with cardiac defects are 35.9\%, 42.5\%, 42.3\%, 43.6\%, 47.0\%, 46.2\% for one-angle, 3D-CNN, Dual-3D CNN, TIP-Net, and four-angle image volumes, respectively. Images reconstructed by TIP-Net present overall higher defect contrast and the measured defect size is closest to the four-angle images. 

For normal porcine and physical phantom studies without cardiac defect, these numbers are 16.0\%, 12.0\%, 14.7\%, 11.2\%, 11.5\%, and 10.1\%. All neural networks showed improved uniformity in the myocardium with lower measured defect size. Both transformer-based methods, TIP-Net and SSTrans-3D, showed better defect quantification than other methods on these normal subjects.

\begin{table}
\scriptsize
\centering
\caption{Quantitative values for porcine and phantom results obtained using different methods ($\mathrm{MEAN}\pm \mathrm{STD}$). Images reconstructed using four-angle data were used as the reference. Best values are marked in bold. $p<0.05$ observed between all groups.}\label{tab1}
\scalebox{0.93}{
\begin{tabular}{|c|c|c|c|c|c|c|}
\hline
\textbf{Evaluation} & \textbf{One-angle} & \textbf{3D-CNN} & \textbf{Dual-3D-CNN} & \textbf{SSTrans-3D} & \textbf{TIP-Net} & \textbf{Four-angle}\Tstrut\Bstrut\\
\hline
\textbf{SSIM$\uparrow$ } & $0.945\pm0.009$ & $0.951\pm0.008$ & $0.950\pm0.008$ & $0.947\pm0.009$ & \pmb{$0.953\pm0.008$} & \diagbox{}{}\Tstrut\Bstrut\\
\hline
\textbf{PSNR$\uparrow$} & $35.764\pm3.546$ & $36.763\pm3.300$ & $36.787\pm3.208$ & $36.613\pm3.509$ & \pmb{$38.114\pm2.876$} & \diagbox{}{} \Tstrut\Bstrut\\
\hline
\textbf{RMSE$\downarrow$} & $0.020\pm0.005$ & $0.0181\pm0.005$ & $0.0185\pm0.005$ & $0.0194\pm0.005$ & \pmb{$0.0179\pm0.005$} & \diagbox{}{} \Tstrut\Bstrut\\
\hline
\textbf{MBP$\uparrow$} & $5.16\pm1.83$ & $10.73\pm2.98$ & $10.21\pm3.17$ & $11.10\pm2.78$ & \pmb{$12.73\pm3.73$} & $13.63\pm3.91$ \Tstrut\Bstrut\\
\hline
\end{tabular}}
\end{table}

\subsection{Human results}
\begin{figure*}[!ht]
\centering
\makebox[\textwidth][c]{\includegraphics[width=\textwidth]{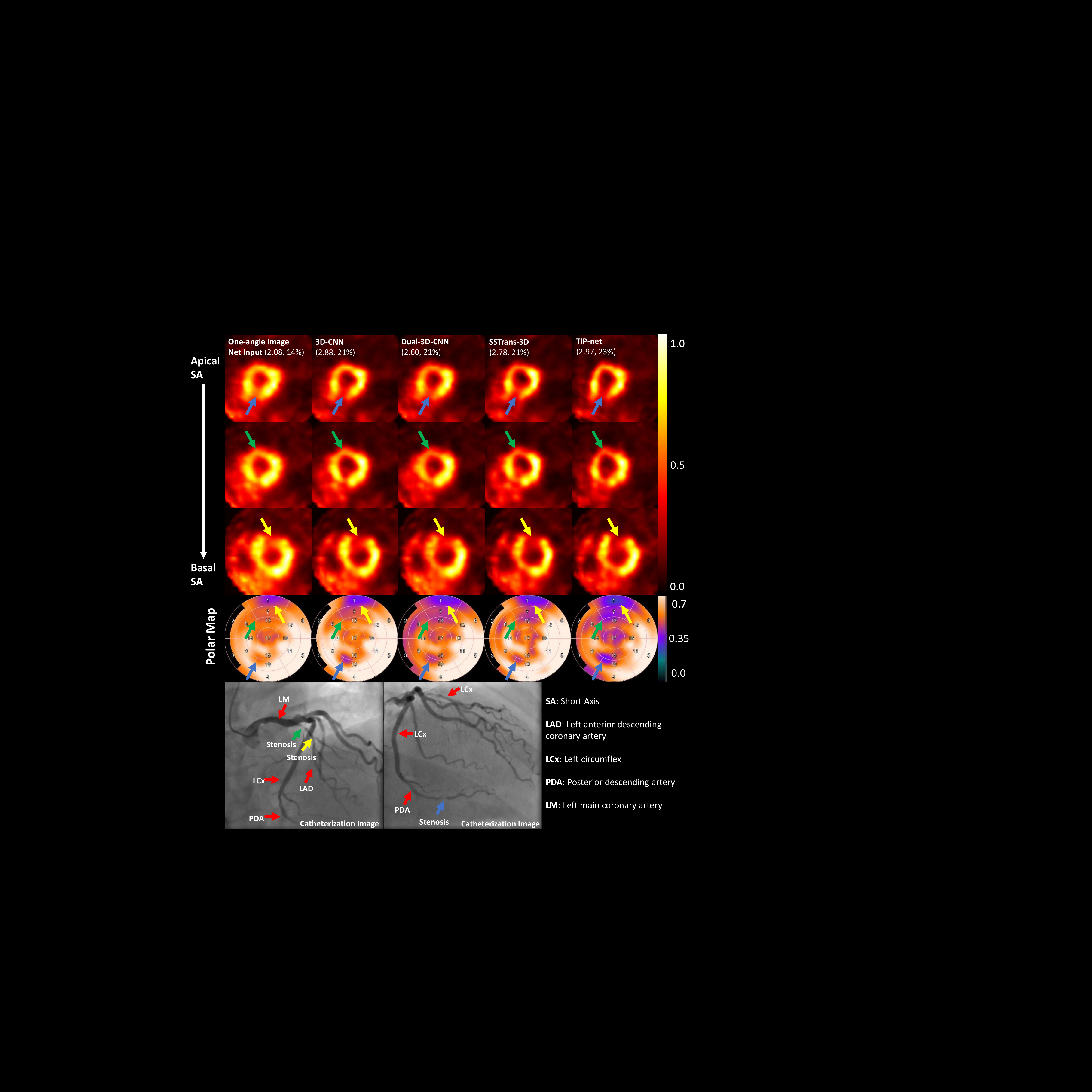}}
\centering
\caption{Results from a sample stress human study. Red arrows point to different coronary vessels. Non-red arrows with the same color point to the same defect/stenosis in the heart. Numbers in the parentheses are the MBP ratios and the defect size (\%LV).}
\label{fig:2}
\end{figure*}

Chosen by a nuclear cardiologist, results from one representative human study with multiple cardiac defects are presented in Fig. \ref{fig:2}. Note that there was no four-angle data for human studies. Based on clinical interpretations from the nuclear cardiologist and Cath results, this patient had multiple perfusion defects in the apical (blue arrows in Fig. \ref{fig:2}) and basal (green and yellow arrows in Fig. \ref{fig:2}) regions. As presented in Fig. \ref{fig:2}, all the networks produced images with better defect contrasts. The proposed TIP-Net further improved the defect contrast, which is favorable and could make the clinical interpretations more confident.

\begin{figure*}[!ht]
\centering
\makebox[\textwidth][c]{\includegraphics[width=0.9\textwidth]{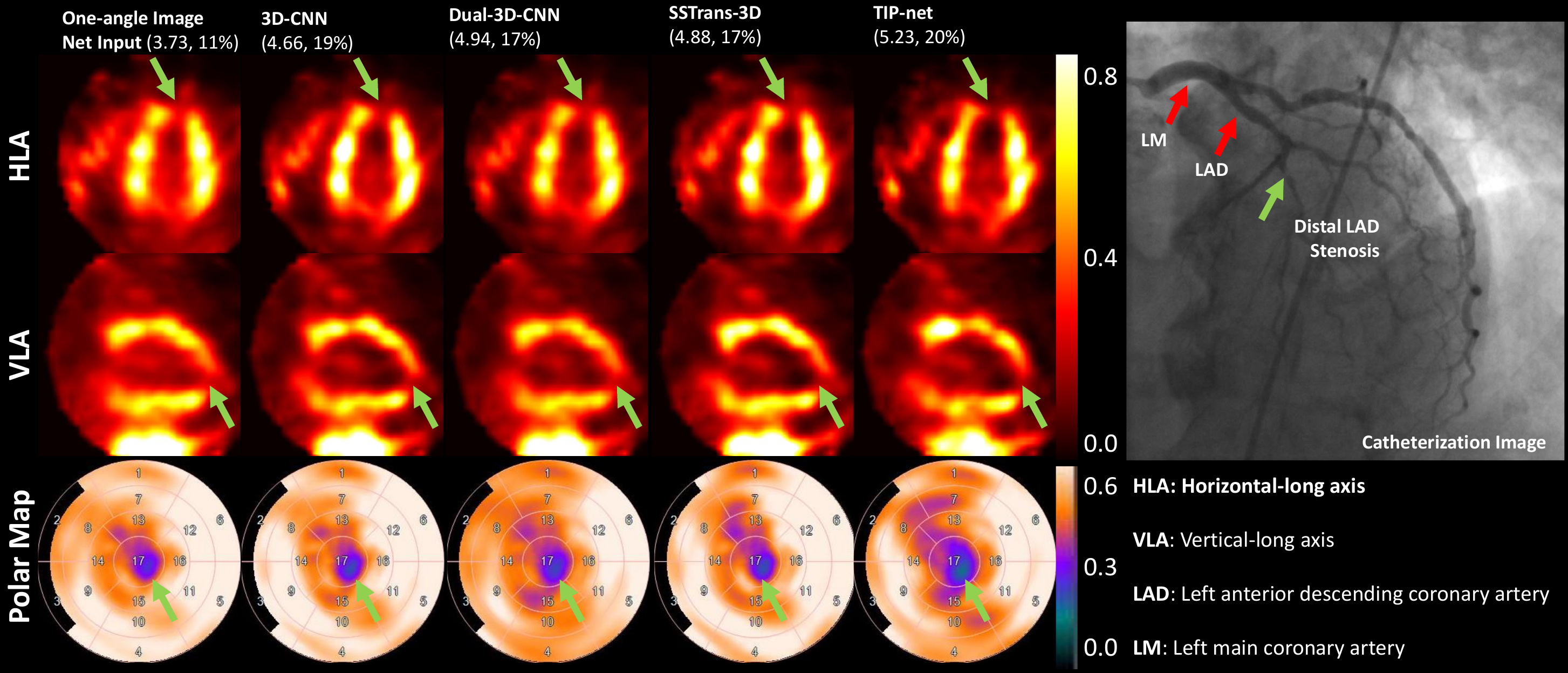}}
\centering
\caption{Results from a sample stress human study. Red arrows point to different coronary vessels. Light-green arrows point to the same defect/stenosis in the heart. Numbers in the parentheses are the MBP ratios and the defect size (\%LV).}
\label{fig:3}
\end{figure*}

Another human study was selected and presented in Fig. \ref{fig:3}. This patient had stenosis in the distal left anterior descending coronary artery, leading to a medium-sized defect in the entire apical region (light-green arrows). As validated by the Cath images, polar map, and interpretation from a nuclear cardiologist, TIP-Net produced images with higher apical defect contrast compared with other methods, potentially leading to a more confident diagnostic decision.

The average MBP ratios on human studies for one-angle, 3D-CNN, Dual-3D-CNN, SSTrans-3D, and TIP-Net images are $3.13\pm0.62$, $4.08\pm0.83$, $4.31\pm1.07$, $4.17\pm0.99$, and $4.62\pm1.29$, respectively ($\mathrm{MEAN}\pm \mathrm{STD}$). The higher ratios in images produced by TIP-Net typically indicate higher image resolution.

14 patients in the testing data have cardiac defects, according to diagnostic results. The average defect size measurements for these patients are 16.8\%, 22.6\%, 21.0\%, 22.4\%, and 23.6\% for one-angle, 3D-CNN, Dual-3D-CNN, SSTrans-3D, and TIP-Net results. The higher measured defect size of the TIP-Net indicates that the proposed TIP-Net produced images with higher defect contrast. 

For the other 6 patients without cardiac defects, these numbers are 11.5\%, 12.8\%, 13.8\%, 12.4\%, and 11.8\%. These numbers show that TIP-Net did not introduce undesirable noise in the myocardium, maintaining the overall uniformity for normal patients. However, other deep learning methods tended to introduce non-uniformity in these normal patients and increased the defect size.

\subsection{Intermediate network output}
To further show the effectiveness of P-net in the overall TIP-Net design, outputs from P-net (images reconstructed by the network directly from projections) are presented in Fig. S2. The presented results demonstrate that the network can learn the mapping between 3D projections to 3D image volumes directly without the iterative reconstruction process. In the porcine study, limited angular sampling introduced streak artifacts in the MLEM-reconstructed one-angle images (yellow arrows in Fig. S2). P-net produced images with fewer few-view artifacts and higher image resolution.

The human study presented in Fig. S2 has an apical defect according to the diagnostic results (blue arrows in Fig. S2). However, this apical defect is barely visible in the one-angle image. P-net produced images with higher resolution and improved defect contrast. Combining both outputs (one-angle images and $\mathrm{IMG_p}$), TIP-Net further enhanced the defect contrast.

\section{Discussion and Conclusion}

We proposed a novel TIP-Net for 3D cardiac SPECT reconstruction. To the best of our knowledge, this work is the first attempt to learn the mapping from 3D realistic projections to 3D image volumes. Previous works in this direction \cite{zhu_image_2018,li_learning_2019,he_radon_2020} were 2D and only considered simulated projections with ideal conditions. 3D realistic projection data have more complex geometry are also affected by other physical factors that may not exist in simulated projections.

The proposed method was tested for myocardial perfusion SPECT imaging. Validated by nuclear cardiologists, diagnostic results, Cath images, and defect size measured by WLCQ, the proposed TIP-Net produced images with higher resolution and higher defect contrast for patients with perfusion defects. For normal patients without perfusion defects, TIP-Net maintained overall uniformity in the myocardium with higher image resolution. Similar performance was observed in porcine and physical phantom studies. 
%
%
%
%
\bibliographystyle{splncs04}
\bibliography{paper1015.bib}

\clearpage

\renewcommand{\thefigure}{S\arabic{figure}}
\renewcommand{\thetable}{S\arabic{table}}  

\begin{center}
\textbf{\large Supplemental Materials}
\end{center}
\setcounter{figure}{0}
\section{Additional Figures}

\begin{figure*}[!ht]
\centering
\makebox[\textwidth][c]{\includegraphics[width=\textwidth]{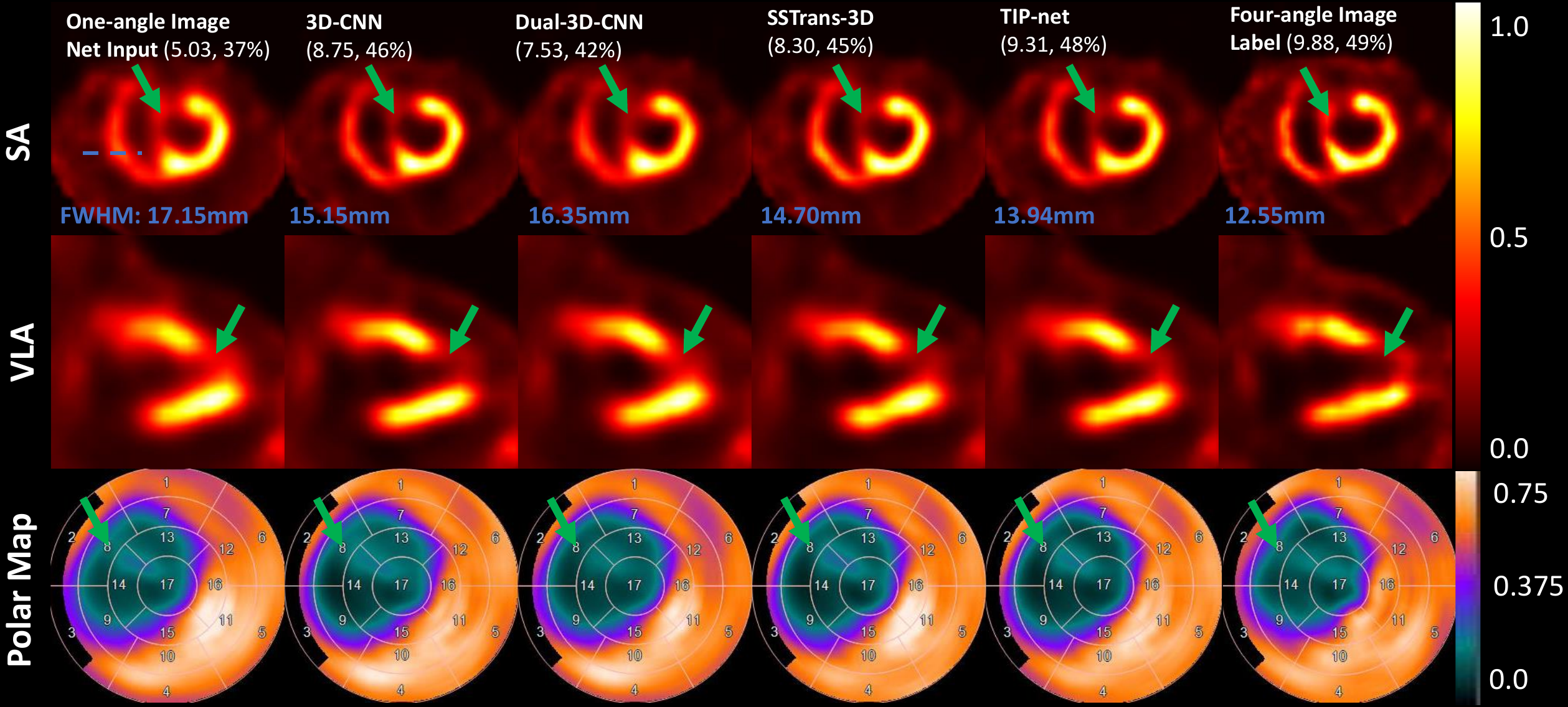}}
\centering
\caption{A sample porcine study. Green arrows point to the post-occlusion reperfusion defect in the heart. Numbers in the parentheses are the MBP ratios and the defect size (\%LV). FWHM values were calculated along the dashed blue lines through the RV myocardium in SA slices. VLA: vertical-long axis. SA: short-axis.}
\label{fig:pig}
\end{figure*}

\begin{figure*}[!ht]
\centering
\makebox[\textwidth][c]{\includegraphics[width=\textwidth]{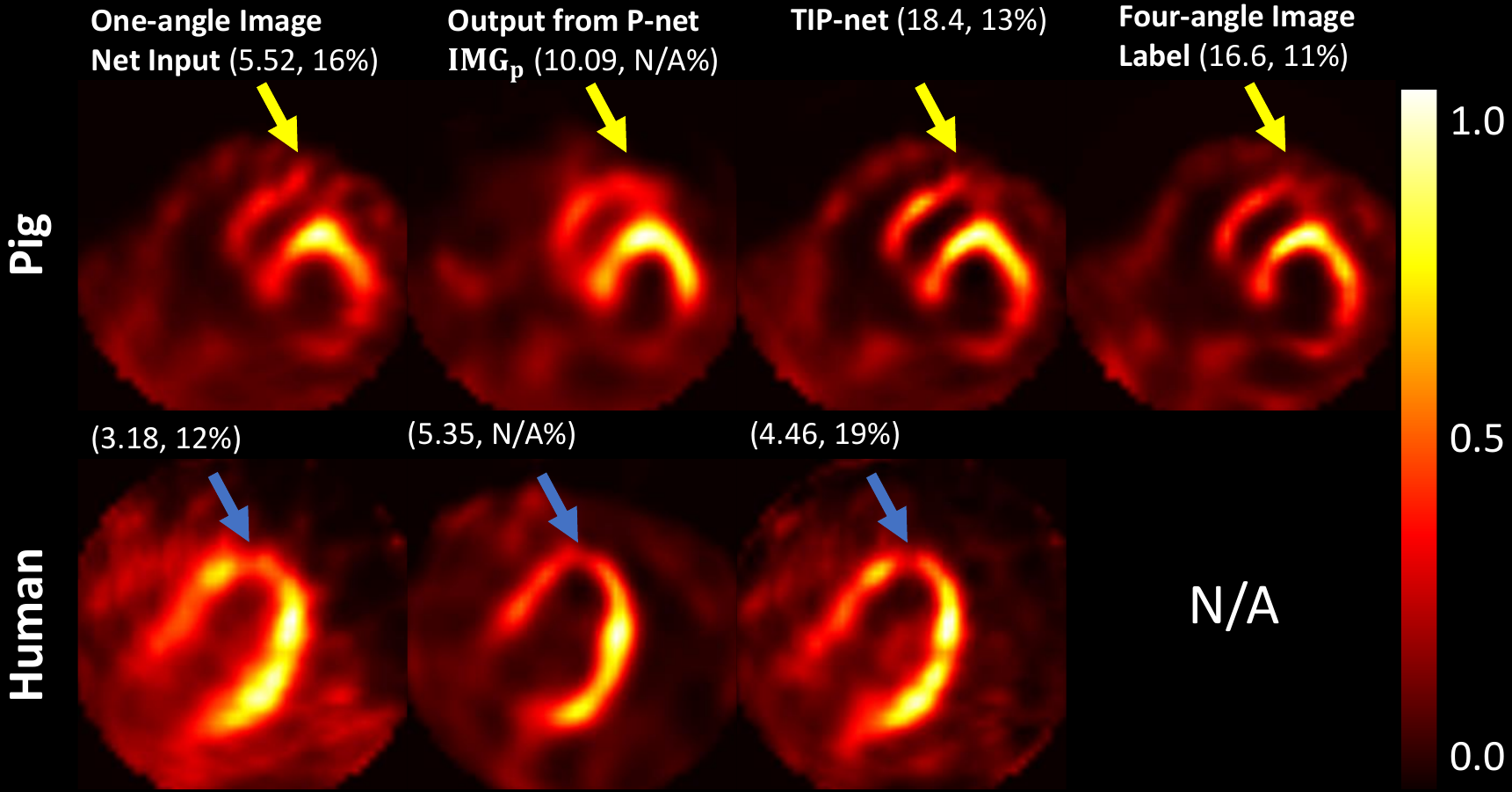}}
\centering
\caption{Network outputs at different stages in TIP-Net. Yellow arrows in pig study point to streak artifacts were introduced due to limited angular sampling in one-angle images. Blue arrows point to the cardiac defect, confirmed by cardiologist, which is barely invisible in the one-angle image. Numbers in the parentheses are the calculated MBP ratios and the defect size (\%LV). Defect size measurements were not available for intermediate network output (i.e., $\mathrm{IMG_p}$).}
\label{fig:4}
\end{figure*}

\end{document}